\documentclass[runningheads]{llncs}
\usepackage{graphicx}
\usepackage[misc]{ifsym} 

\usepackage{url}

\usepackage{hyperref}

\usepackage[table]{xcolor}
\usepackage{pgfplotstable}
\usepackage{booktabs}
\usepackage{adjustbox}
\usepackage[cmex10]{amsmath}
\usepackage{array} 

\DeclareMathOperator*{\argmax}{argmax} 

\usepackage{multirow}
\usepackage{amssymb}

\pgfplotsset{compat=1.14}
\usepgfplotslibrary{colormaps}

\newcolumntype{+}{>{\global\let\currentrowstyle\relax}}
\newcolumntype{^}{>{\currentrowstyle}}
\newcommand{\rowstyle}[1]{\gdef\currentrowstyle{#1}%
#1\ignorespaces
}

\newcolumntype{C}{>{\centering\arraybackslash}p{8mm}} 

\usepackage{array}
\newcommand{\PreserveBackslash}[1]{\let\temp=\\#1\let\\=\temp}
\newcolumntype{C}[1]{>{\PreserveBackslash\centering}p{#1}}
\newcolumntype{R}[1]{>{\PreserveBackslash\raggedleft}p{#1}}
\newcolumntype{L}[1]{>{\PreserveBackslash\raggedright}p{#1}}

\pgfplotstableset{
    /color cells/min/.initial=0,
    /color cells/max/.initial=1000,
    /color cells/textcolor/.initial=,
    color cells/.code={%
        \pgfqkeys{/color cells}{#1}%
        \pgfkeysalso{%
            postproc cell content/.code={%
                \begingroup
                \pgfkeysgetvalue{/pgfplots/table/@preprocessed cell content}\value
\ifx\value\empty
\endgroup
\else
                \pgfmathfloatparsenumber{\value}%
                \pgfmathfloattofixed{\pgfmathresult}%
                \let\value=\pgfmathresult
                \pgfplotscolormapaccess
                    [\pgfkeysvalueof{/color cells/min}:\pgfkeysvalueof{/color cells/max}]%
                    {\value}%
                    {\pgfkeysvalueof{/pgfplots/colormap name}}%
                \pgfkeysgetvalue{/pgfplots/table/@cell content}\typesetvalue
                \pgfkeysgetvalue{/color cells/textcolor}\textcolorvalue
                \toks0=\expandafter{\typesetvalue}%
                \xdef\temp{%
                    \noexpand\pgfkeysalso{%
                        @cell content={%
                            \noexpand\cellcolor[rgb]{\pgfmathresult}%
                            \noexpand\definecolor{mapped color}{rgb}{\pgfmathresult}%
                            \ifx\textcolorvalue\empty
                            \else
                                \noexpand\color{\textcolorvalue}%
                            \fi
                            \the\toks0 %
                        }%
                    }%
                }%
                \endgroup
                \temp
\fi
            }%
        }%
    }
}

\begin{document}
\title{Biased RSA private keys: Origin attribution of GCD-factorable keys\thanks{Full details, datasets and paper supplementary material can be found at \url{https://crocs.fi.muni.cz/papers/privrsa_esorics20}}}
%
%
\author{%
Adam Janovsky\inst{1,2}\textsuperscript{(\Letter)} \and
Matus Nemec\inst{3} \and
Petr Svenda\inst{1} \and
Peter Sekan\inst{1} \and
Vashek Matyas\inst{1}
}
\institute{
Masaryk University, Czech Republic \\ \Letter~\email{adamjanovsky@mail.muni.cz}
\and
Invasys, Czech Republic
\and
Link\"oping University, Sweden
}
\authorrunning{Janovsky et al.}
\maketitle

\newcommand{\numlibs}{70 }
\newcommand{\numlibsversions}{70 }
\newcommand{\numgroups}{26 }
\newcommand{\numsimplefeatures}{35 }
\newcommand{\bestaccuracysinglekey}{47}
\newcommand{\bestfeatures}{\textbf{5p\_5q\_blum\_mod\_roca} } 

\begin{abstract}
In 2016, Švenda et al. (USENIX 2016, The Million-key Question) reported that the implementation choices in cryptographic libraries allow for qualified guessing about the origin of public RSA keys.
We extend the technique to two new scenarios when not only public but also private keys are available for the origin attribution -- analysis of a source of GCD-factorable keys in IPv4-wide TLS scans and forensic investigation of an unknown source. We learn several representatives of the bias from the private keys to train a model on more than 150 million keys collected from 70 cryptographic libraries, hardware security modules and cryptographic smartcards. Our model not only doubles the number of distinguishable groups of libraries (compared to public keys from Švenda et al.) but also improves more than twice in accuracy w.r.t. random guessing when a single key is classified. For a forensic scenario where at least 10 keys from the same source are available, the correct origin library is correctly identified with average accuracy of 89\% compared to 4\% accuracy of a random guess. The technique was also used to identify libraries producing GCD-factorable TLS keys, showing that only three groups are the probable suspects. 


\end{abstract}

\section{Introduction}
The ability to attribute a cryptographic key to the library it was generated with is a valuable asset providing direct insight into cryptographic practices. The slight bias found specifically in the primes of RSA private keys generated by the OpenSSL library~\cite{heninger2012mining} allowed to track down the devices responsible for keys found in TLS IPv4-wide scans that were in fact factorable by distributed GCD algorithm. Further work \cite{1mrsa} made the method generic and showed that many other libraries produce biased keys allowing for the origin attribution. As a result, both separate keys, as well as large datasets, could be analyzed for their origin libraries. 
The first-ever explicit measurement of cryptographic library popularity was introduced in~\cite{2017-acsac-nemec}, showing the increasing dominance of the OpenSSL library on the market.
Furthermore, very uncommon characteristics of the library used by Infineon smartcards allowed for their entirely accurate classification. Importantly, this led to a discovery that the library is, in fact, producing practically factorable keys~\cite{roca-2017-ccs-nemec}.
Consequently, more than 20 million of eID certificates with vulnerable keys were revoked just in Europe alone. The same method allowed to identify keys originating from unexpected sources in Estonian eIDs. Eventually, the unexpected keys were shown to be injected from outside instead of being generated on-chip as mandated by the institutional policy~\cite{parsovs_usenix}.

While properties of RSA primes were analyzed to understand the bias detected in public keys, no previous work addressed the origin attribution problem \emph{with} the knowledge of private keys. The reason may sound understandable -- while the public keys are readily available in most usage domains, the private keys shall be kept secret, therefore unavailable for such scrutiny. Yet there are at least two important scenarios for their analysis: 1) Tracking sources of GCD-factorable keys from large TLS scans and 2) a forensic identification of black-box devices with the capability to export private keys (e.g., unknown smartcard, remote key generation service, or in-house investigation of cryptographic services). The mentioned case of unexpected keys in Estonian eIDs \cite{parsovs_usenix} is a practical example of a forensic scenario, but with the use of public keys only. The analysis based on private keys can spot even a smaller deviance from the expected origin as the bias is observed closer to the place of its inception. 
This work aims to fill this gap in knowledge by a careful examination of both scenarios.

We first provide a solid coverage of RSA key sources used in the wild by expanding upon the dataset first released in~\cite{1mrsa}. During our work, we more than doubled the number of keys in the dataset, gathered from over \numlibs distinct cryptographic software libraries, smartcards, and hardware security modules (HSMs). Benefiting from 158.8 million keys, we study the bias affecting the primes $p$ and $q$. 
We transform known biased features of public keys to their private key analogues and evaluate how they cluster sources of RSA keys into groups. We use the features in multiple variants of Bayes classifier that are trained on 157 million keys.
 Subsequently, we evaluate the performance of our classifiers on further 1.8 million keys isolated from the whole dataset. By doing so, we establish the reliability results for the forensic case of use, when keys from a black-box system are under scrutiny. On average, when looking  at just a single key, our best model is able to correctly classify \bestaccuracysinglekey\% of cases when all libraries are considered and 64.6\% keys when the specific sub-domain of smartcards is considered. These results allow for much more precise classification compared to the scenario when only public keys are available.
 
 Finally, we use the best-performing classification method to analyze the dataset of GCD-factorable RSA keys from the IPv4-wide TLS scan collected by Rapid7~\cite{rapid7}.
\newpage
\noindent The main contributions of this paper are:
\begin{itemize}
\item A systematic mapping of biased features of RSA keys evaluated on a more exhaustive set of cryptographic libraries, described in Section \ref{sec:rsa_bias}. The dataset (made publicly available for other researchers) lead to \numgroups total groups of libraries distinguishable based on the features extracted from the value of RSA private key(s). 
\item Detailed evaluation of the dataset on Bayes classifiers in Section \ref{sec:accuracy} with an average accuracy above $\bestaccuracysinglekey\%$  where only a single key is available, and almost $90\%$ when ten keys are available.
\item An analysis of the narrow domain of cryptographic smartcards and libraries used for TLS results in an even higher accuracy, as shown in Section \ref{sec:domain_specific}.
\item Practical analysis of real-world sources of GCD-factorable RSA keys from public TLS servers obtained from internet-wide scans in Section \ref{sect:world_results}.
\end{itemize}

The paper roadmap has been partly outlined above, Section~\ref{sect:discusussion} then shows related work and Section~\ref{sect:conclusions} concludes our paper.

\section{Bias in RSA keys}
\label{sec:rsa_bias}
Various design and implementation decisions in the algorithms for generating RSA keys influence the distributions of produced RSA keys. A specific type of bias was used to identify OpenSSL as the origin of a group of private keys \cite{Mironov}. 
Systematic studies of a wide range of libraries \cite{1mrsa,2017-acsac-nemec} described more reasons for biases in RSA keys in a surprising number of libraries. In the majority of cases, the bias was not strong enough to help factor the keys more efficiently. 
Previous research \cite{1mrsa} identified multiple sources of bias that our observations from a large dataset of private RSA keys confirm:

\begin{enumerate}
\item[\textbf{1.}] \textbf{Performance optimizations}, e.g., most significant bits of primes set to a fixed value to obtain RSA moduli of a defined length.

\item[\textbf{2.}] \textbf{Type of primes}: probable, strong, and provable primes:
    \begin{itemize}
        \item For probable primes, whether candidate values for primes are chosen randomly or a single starting value is incremented until a prime is found.
        \item When generating candidates for probable primes, small factors are avoided in the value of $p-1$ by multiple implementations without explaining.
        \item Blum integers are sometimes used for RSA moduli -- both RSA primes are congruent to 3 modulo 4.
        \item For strong primes, the size of the auxiliary prime factors of $p-1$ and $p+1$ is biased.
        \item For provable primes, the recursive algorithm can create new primes of double to triple the binary length of a given prime; usually one version of the algorithm is chosen.
    \end{itemize}
    
\item[\textbf{3.}] \textbf{Ordering of primes}: are the RSA primes in private key ordered by size?

\item[\textbf{4.}] \textbf{Proprietary algorithms}, e.g., the well-documented case of Infineon fast prime key generation algorithm  \cite{roca-2017-ccs-nemec}.

\item[\textbf{5.}] \textbf{Bias in the output of a PRNG}: often observable only from a large number of keys from the same source;

\item[\textbf{6.}] \textbf{Natural properties of primes} that do not depend on the implementation.

\end{enumerate}

\subsection{Dataset of RSA keys}
We collected, analyzed, and published the largest dataset of RSA keys with a known origin from \numlibsversions libraries (43 open-source libraries, 5 black-box libraries, 3 HSMs, 19 smartcards). We both expanded the datasets from previous work \cite{1mrsa,2017-acsac-nemec} and generated new keys from additional libraries for the sake of this study. We processed the keys to a unified format and made them publicly available. Where possible, we analyzed the source code of the cryptographic library to identify the basic properties of key generation according to the list above.

We are primarily interested in 2048-bit keys, what is the most commonly used key length for RSA. As in previous studies \cite{1mrsa,2017-acsac-nemec}, we also generate shorter keys (512 and 1024 bits) to speed up the process, while verifying that the chosen biased features are not influenced by the key size. This makes the keys of different sizes interchangeable for the sake of our study. We assume that repeatedly running the key generation locally approximates the distributed behaviour of many instances of the same library. This model is supported by the measurements taken in \cite{2017-acsac-nemec} where distributions of keys collected from the Internet exhibited the same biases as locally generated keys.

\subsection{Choice of relevant biased features}
\label{sec:features}

We extended the features used in previous work on public keys to their equivalent properties of private keys:

\noindent \textbf{Feature `5p and 5q'}: Instead of the most significant bits of the modulus, we use five most significant bits of the primes p and q. The modulus is defined by the primes, and the primes naturally provide more information. We chose 5 bits based on a frequency analysis of high bits. Further bits are typically not biased and reducing the size of this feature prevents an exponential growth of the feature space.

\noindent \textbf{Feature `blum'}: We replaced the feature of second least significant bit of the modulus by the detection of Blum integers. Blum integers can be directly identified using the two prime factors. When only the modulus is available, we can rule out the usage of Blum integers, but not confirm it.

\noindent \textbf{Feature `mod'}: Previous work used the result of modulus modulo 3. It was known that primes can be biased modulo small primes (due to avoiding small factors of $p-1$ and $q-1$). The authors only used the value 3, because it is possible to rule out that 3 is being avoided as a factor of $p-1$, when the modulus equals 2 modulo 3 \cite{1mrsa}. It is not possible to rule out higher factors from just a single modulus.
With the access to the primes we can directly check for this bias for all factors. We detected four categories of such bias, each avoiding all small odd prime factors up to a threshold. We use these categories directly by looking at small odd divisors of $p-1$ and $q-1$ and note if none were detected: 1) up to 17863, 2) up to 251, 3) up to 5, 4) none -- at least one value is divisible by 3.

\noindent \textbf{Feature `roca'}: We use a specific fingerprint of factorable Infineon keys published in \cite{roca-2017-ccs-nemec}.

\subsection{Clustering of sources into groups}
\label{sec:clustering}

\begin{figure}[t]
    \centering
    \includegraphics[width=1\textwidth]{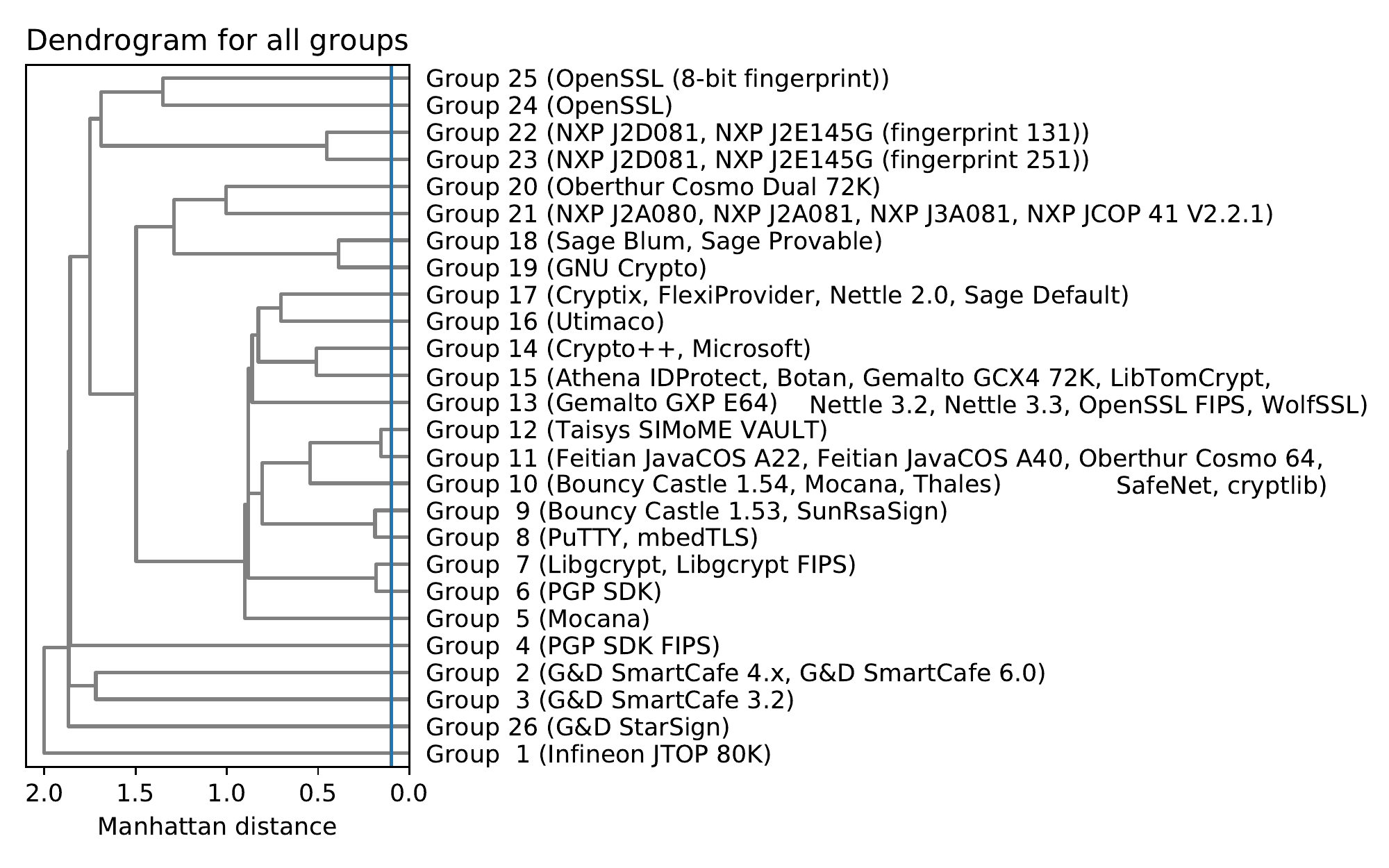}
\caption{How the keys from various libraries differ can be depicted by a dendrogram. It tells us, w.r.t. our feature set, how far from each other the probability distributions of the sources are. We can then hierarchically cluster the sources into groups that produce similar keys. The blue line at 0.085 highlights the threshold of differentiating between two sources/groups. This threshold yields \numgroups groups using our feature set.}
\label{fig:private_key_dendogram}
\end{figure}

Since it is impossible to distinguish sources that produce identically distributed keys, we introduce a process of clustering to merge similar sources into groups. We cluster two sources together if they appear to be using identical algorithms based on the observation of the key distributions. We measure the difference in the distributions using the Manhattan distance\footnote{We experimented with Euclidean distance and fractional norms. While Euclidean distance is a proper metric, our experiments showed that it is more sensitive to the noise in the data, creating separable groups out of sources that share the same key generation algorithms. On the other hand, fractional norms did not highlight differences between sources that provably differ in the key generation process.}. The absolute values of the distances depend on the actual distributions of the features. Large distances correlate with significant differences in the implementations. Note, that very small observed distances may be only the result of noise in the distributions instead of a real difference, e.g., due to a smaller number of keys available.

We attempt to place the clustering threshold as low as possible, maximizing the number of meaningful groups. If we are not able to explain why two clusters are separated based on the study of the algorithms and distributions of the features, the threshold needs to be moved higher to join these clusters. We worked with distributions that assume all features correlated (as in \cite{1mrsa}). 

The resulting classification groups and the dendrogram is shown in Figure \ref{fig:private_key_dendogram}. 
We placed the threshold value at 0.085. By moving it higher than to 0.154, we would lose the ability to distinguish groups 11 and 12. It would be possible to further split group 14, as there is a slight difference in the prime selection intervals used by Crypto++ and Microsoft \cite{1mrsa}. However, the difference manifests less than the level of noise in other sources, requiring the threshold to be put at 0.052, what would create several false groups. We use the same clustering throughout the paper, although the value of the threshold would change when the features change. Note that different versions of the same library may fall into different groups, mostly because of the algorithm changes between these versions. This, for instance, is the case of the Bouncy Castle 1.53, and 1.54. 

\section{Model selection and evaluation}
\label{sec:accuracy}
How accurately we can classify the keys depends on several factors, most notably on: the libraries included in the training set, number of keys available for classification, features extracted from the classified keys, and on the classification model. In this section, we focus on the last factor.

\subsection{Model selection}
\label{sect:methodOverview}

As generating the RSA keys is internally a stochastic process, we choose the family of probabilistic models to address the source attribution problem. Since there is no strong motivation for complex machine learning models, we utilize simple classifiers. More sophisticated classifiers could be built based on our findings when the goal is to reach higher accuracy or to more finely discriminate sources within a group. The rest of this subsection describes the chosen models.

\textbf{Naïve Bayes classifier.}
The first investigated model is a \emph{naïve Bayes classifier}, called naïve because it assumes that the underlying features are conditionally independent. Using this model, we apply the maximum-likelihood decision rule and predict the label as $\hat{y} = \argmax_y P(X=x \mid y)$. Thanks to the naïve assumption, we may decompose this computation into $\hat{y} = \argmax_y \prod_{i=1}^n P(x_i \mid y)$ for the feature vector $x = (x_1, \dots, x_n)$.

\textbf{Bayes classifier.}
\label{sec:classicBayes}
We continue to develop the approach originally used in~\cite{1mrsa} that used the \emph{Bayes classifier} without the naïve assumption. Several reasons motivate this. First, it allows to evaluate how much the naïve Bayes model suffers from the violated independence assumption (on this specific dataset). Secondly, it enables us to access more precise probability estimates that are needed to classify real-world GCD-factorable keys. Additionally, we can directly compare the classification accuracy of private keys with the case of the public keys from~\cite{1mrsa}. However, one of the main drawbacks of the Bayes classifier is that it requires exponentially more data with the growing number of features. Therefore, when striving for high accuracy achievable by further feature engineering, one should consider the naïve Bayes instead. 

\textbf{Naïve Bayes classifier with cross-features.}
\label{sec:naiveBayesCombined}
The third investigated option is the naïve Bayes classifier, but we merged selected features that are known to be correlated into a single feature. In particular, we merged the features of the most significant bits (of $p,q$) into a single cross-feature. Subsequently, the naïve Bayes approach is used. This enables us to evaluate whether merging clearly interdependent features into one will affect the performance of naïve Bayes classifier w.r.t. this specific dataset.

\subsection{Model evaluation} 

\subsubsection{Methodology of classification and metrics.}

Our training dataset contains 157 million keys and the test set contains 1.8 million keys. We derived the test set by discarding 10 thousand keys of each source from the complete dataset before clustering. This assures that each group has the test set with at least 10 thousand keys. Accordingly, since the groups differ in the number of sources involved, the resulting test dataset is imbalanced. For this reason, we employ the metrics of precision and recall when possible. However, we represent the model performance by accuracy measure in the tables and in more complex classification scenarios.

For \emph{group X}, the precision can be understood as a fraction of correctly classified keys from \emph{group X} divided by the number of keys that were marked as \emph{group X} by our classifier. Similarly, the recall is a fraction of correctly classified keys from \emph{group X} divided by a total number of keys from \emph{group X}~\cite{mlbook}. We also evaluate the performance of the models under the assumption that the user has a \emph{batch} of several keys from the same source at hand. This scenario can arise, e.g., when a security audit is run in an organization and all keys are being tested. Furthermore, to react to some often misclassified groups, we additionally provide the answer ``this key originates from \emph{group X} or \emph{group Y}'' to the user (and we evaluate the confidence of these answers).

\subsubsection{Comparison of the models.}

\begin{table}[b]
    \centering
    \begin{tabular}{|l|c|c|}
        \hline
        Model & Avg. precision & Avg. recall \\  [1mm]
        \hline        
        Bayes classifier & $43.2\%$ & $47.6\%$ \\ 
        Naïve Bayes classifier & $40.9\%$ & $46.2\%$ \\ 
        Cross-feature naïve B. & $41.7\%$ & $47.6\%$ \\
        \hline
       \end{tabular}
       \vspace{4mm}
    \caption{Performance comparison of different models on the dataset with all libraries. Note that the precision of a random guess classifier is $3.8$\% when considering 26 groups.}
    \label{table_model_comparision}
\end{table}

The overall comparison of all three models can be seen in Table~\ref{table_model_comparision}. If the precision for some group is undefined, i.e., no key is allegedly originating from this group, we say that the precision is 0.
We evaluate the naïve Bayes classifier on the same features that were used for Bayes classifier to measure how much classification performance is lost by introducing the feature independence assumption. A typical example of interdependent features is that the most significant bits of primes $p$ and $q$ are intentionally correlated to preserve the expected length of the resulting modulus $n$. Pleasantly, the observed precision (recall) decrease is only $2.3\%$ ($1.4\%$) when compared to the Bayes classifier. Accordingly, this suggests that a larger number of different features than usable with the Bayes classifier (due to exponential growth in complexity) can be considered when the naïve Bayes classifier is used. As a result, further improvement of the performance might be achieved, despite ignoring the dependencies among features. Overall, the Bayes classifier shows the best results. When a single key is classified, the average success rate for the 26 groups is captured by precision of $43.2\%$ and a recall of $47.6\%$. Still, there is a wide variance between the performance in specific groups. A detailed table of results together with a discussion is presented in Appendix~\ref{subsec:discussion_results}.

\section{Classification with prior information}
\label{sec:domain_specific}

Section~\ref{sec:rsa_bias} outlined the process of choosing a threshold value that determines the critical distance for distinguishing between distinct groups. Inevitably, the same threshold value directly influences the number of groups after the clustering task. As such, the threshold introduces a trade-off between the model performance and the number of discriminated groups. 
The smaller the difference between group distributions is, the more they are similar, and the model performance is lower as more misclassification errors occur. The objective of this section is to examine the classification scenario when some prior knowledge is available to the analyst, limiting the origin of keys to only a subset of all libraries or increase the likelihood of some. Since Section~\ref{sec:accuracy} showed that the Bayes classifier provides the best performance, this chapter considers only this model. 

Prior knowledge can be introduced into the classification process in multiple ways, e.g., by using a prior probability vector that considers some groups more prevalent. We also note that the measurement method of \cite{2017-acsac-nemec} can be used to obtain such prior information, but a relatively large dataset (around $10^5$ private keys) is required that may not be available. Our work, therefore, considers a different setting when some sources of the keys are ruled-out \textit{before} the classifier is constructed. Such scenario arises e.g., when the analyst knows that the scrutinized keys were generated in an unknown cryptographic smartcard. In such case, HSMs and other sources of keys can thus be omitted from the model altogether what will arguably increase the performance of the classification process. Another example is leaving out libraries that were released after the classified data sample was collected.

We present the classification performance results for three scenarios with a limited number of sources -- 1) cryptographic smartcards (Section \ref{sect:smartcard_domain}), 2) sources likely to be used in the TLS domain (Section \ref{sect:tls_domain}) and 3) a specific case of GCD-factorable keys from the TLS domain, where only one out of two primes can be used for classification (see Section \ref{sect:factorable_tls_domain} for more details). The comparison of models for these scenarios can be seen in Table~\ref{table_domain_comparision}. 

To compute these models we first, discard the sources that cannot be the origin of the examined keys according to the prior knowledge of the domain (e.g., smartcards are not expected in TLS). Next, we re-compute the clustering task to obtain fewer groups than on the dataset with all libraries. Finally, we compute the classification tables for the reduced domain and evaluate the performance.

\begin{table}[t]
    \centering
    \begin{tabular}{|l|c|c|c|}
        \hline
        Dataset & Avg. precision & Avg. recall & Random guess (baseline) \\[1mm]
        \hline
        All libraries & $43.2\%$ & $47.6\%$  & $3.8$\% \\ 
        Smartcards domain& $61.9\%$ & $64.6\%$ &  $8.3$\% \\ 
        TLS domain & $45.5\%$ & $42.2\%$ & $7.7$\% \\
        Single-prime TLS domain & $28.8\%$ & $36.2\%$ & $11.1$\% \\
        \hline
       \end{tabular}
       \vspace{4mm}
    \caption{Bayes classifier performance on three analyzed partitionings of the dataset -- complete dataset with all libraries (\emph{All libraries}), smartcards only (\emph{Smartcards domain}), libraries and HSMs expected to be used for TLS (\emph{TLS domain}) and specific subset of TLS domain where only single prime is available due to the nature of results obtained by GCD factorization method (\emph{Single-prime TLS domain}). Comparison with the  random guess as a baseline is provided (here, accuracy equals precision and recall).}
    \label{table_domain_comparision}
\end{table}

\subsection{Performance in the smartcards domain}
\label{sect:smartcard_domain}

\begin{figure}
    \centering
    \includegraphics[width=0.9\textwidth]{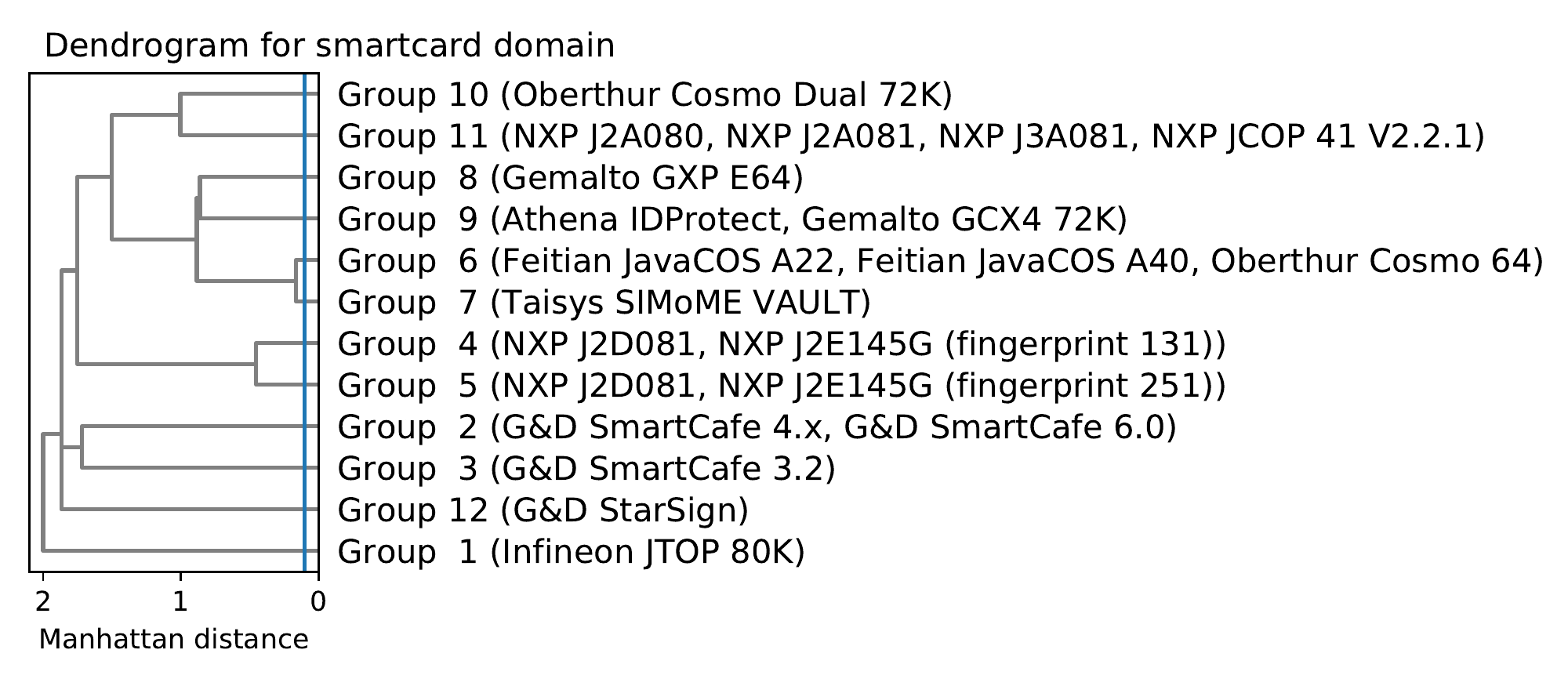}
\caption{The clustering of smartcard sources yields 12 separate groups.}
\label{fig:sc_dendrogram}
\end{figure}

\begin{figure}
    \centering
    \includegraphics[width=0.7\columnwidth]{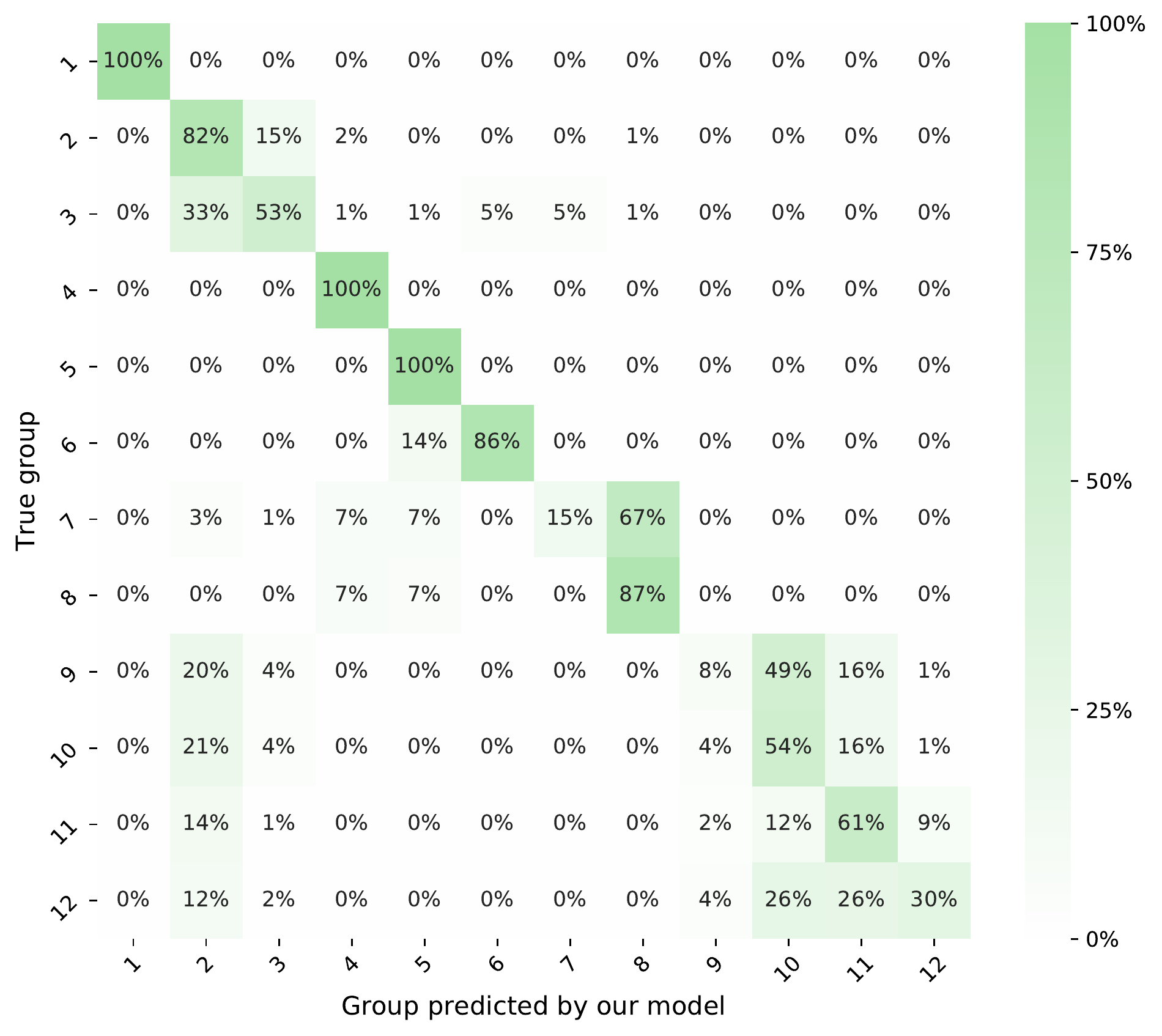}
\caption{The confusion matrix for the classifier of a single private key generated in the smartcards domain. A given row corresponds to a vector of observed relative frequencies with which keys generated by a specific group (True group) are misclassified as generated by other groups (Group predicted by our model). For example, group 1 and group 2 have no misclassifications (high accuracy), while keys of group 3 are in $33\%$ cases misclassified as keys from group 2. On average, we achieve $64.6$\% accuracy. The darker the cell is, the higher number it contains. This holds for all figures in this paper.}\label{fig:confusion_matrix}
\end{figure}

The clustering task in the smartcards domain yields 12 recognizable groups for 19 different smartcard models as shown in Figure \ref{fig:sc_dendrogram}.
The training set for this limited domain contains 20.6 million keys, whereas the test set contains 340 thousand keys. On average, $61.9\%$ precision and $64.6\%$ recall is achieved. Moreover, 8 out of 12 groups achieve $>50\%$ precision. Additionally, the classifier exhibits $100\%$ recall on 3 specific groups: a) Infineon smartcards (before 2017 with the ROCA vulnerability \cite{roca-2017-ccs-nemec}), b) G\&D Smartcafe 4.x and 6.0, and c) newer G\&D Smartcafe 7.0. Figure \ref{fig:confusion_matrix} shows so-called confusion matrix where each row corresponds to percentage of keys in an actual group while each column represents percentage of keys in a predicted group.  

As expected, the results represent an improvement when compared to the dataset with all libraries. When one has ten keys of the same card at hand, the expected recall is over $90\%$ on 10 out of 12 groups. The full table of results can be found in the project repository. 

Interestingly, 512- and 1024-bit keys generated by the same NXP J2E145G card (similarly also for NXP J2D081) fall into different groups\footnote{This is an exception to the observation that the selected features behave independently of key length. Otherwise, keys of different length can be used interchangeably.}. The main difference is in the modular fingerprint (avoidance of small factors in $p-1$ and $q-1$). We hypothesize that on-card key generation avoids more small factors for larger keys. Such behaviour was not observed for other libraries but highlights the necessity of collecting different key lengths in the training dataset when one analyzes black-box proprietary devices or closed-source software libraries.

To summarize, the classification of private keys generated by smartcards is very accurate due to the significant differences resulting from the proprietary, embedded implementations among the different vendors. The differences observed likely results from the requirements to have a smaller footprint required by low-resources devices. 

\subsection{Performance in the TLS domain}
\label{sect:tls_domain}

\begin{figure}
    \centering
    \includegraphics[width=0.9\textwidth]{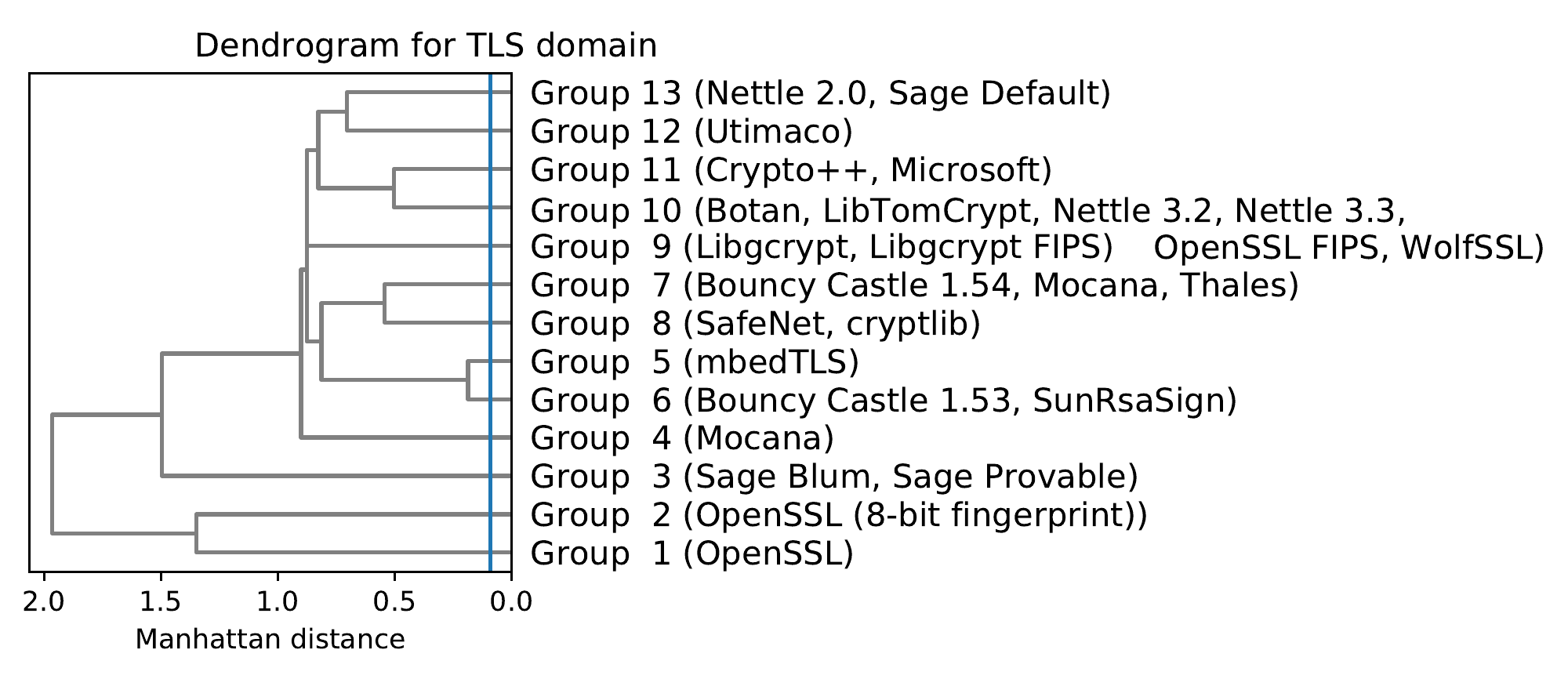}
\caption{The clustering of the sources from the TLS domain yields 13 separate groups.}
\label{fig:tls_dendrogram}
\end{figure}

For the TLS domain, we excluded all the libraries and devices unlikely to be used to generate keys then used by TLS servers. All smartcards are excluded, together with highly outdated or purpose-specific libraries like PGP SDK 4. All hardware security modules (HSMs) are present as they may be used as TLS accelerators or high-security key storage. Summarized, we started with 17 separate cryptographic libraries and HSMs, inspected in a total of 134 versions. The clustering resulted in 13 recognizable groups as shown in Figure \ref{fig:tls_dendrogram}. 

The domain training set contains 121.8 million keys and the test set contains 1.3 million keys. On average, the classifier achieves $45.5\%$ precision and $42.2\%$ recall. The decrease in average recall compared to the full domain may look surprising, but averaging is deceiving in this context. In fact, recall improved for 10 out of 13 groups that are both in the full set and the TLS domain set, with the precision improving for 9 groups. The mean values of the full dataset are being uplifted by a generally better performance of the model outside the TLS domain. Five groups have $>50\%$ precision. OpenSSL (by far the most popular library used by servers for TLS \cite{2017-acsac-nemec}) has $100\%$ recall, making the classification of OpenSSL keys very reliable. Complete results can be found in the project repository. 

To summarize, we correctly classify more keys in a more specific TLS domain than with the full dataset classifier. Additionally, the user can be more confident about the decisions of the TLS-specific classifier. 

\subsection{Performance in the single-prime TLS domain}
\label{sect:factorable_tls_domain}

\begin{table}[b]
    \centering
    \begin{adjustbox}{width=0.75\textwidth, center}
    \begin{tabular}{|@{}c@{}|@{}c@{}|}
		\hline
			{\begin{filecontents*}{accuracy_tables_data/single_prime/groups.csv}
batch_keys
{Group 1}
{Group 2}
{Group 3}
{Group 4}
{Group 5$|$13}
{Group 6}
{Group 7$|$11}
{Group 8$|$9$|$10}
{Group 12}
Average
\end{filecontents*}
\pgfplotstabletypeset[%
                every head row/.style={after row= \hline},
                /pgf/number format/fixed,
                /pgf/number format/precision=1,
                col sep=comma,
                every last row/.style={before row=\hline \rowstyle{\bfseries}},
                columns/batch_keys/.style={column name={Number of primes in a batch}, reset styles,string type,},]{accuracy_tables_data/single_prime/groups.csv}}
			& {\begin{filecontents*}{accuracy_tables_data/single_prime/top_n_1.csv}
n_keys_1,n_keys_10,n_keys_20,n_keys_30,n_keys_100
100.0,100.0,100.0,100.0,100.0
42.8,99.7,100.0,100.0,100.0
78.0,100.0,100.0,100.0,100.0
47.5,90.3,95.8,98.7,100.0
1.8,30.8,43.7,51.8,74.7
5.2,48.9,61.0,64.8,76.7
0.0,67.3,92.3,97.4,100.0
37.9,99.9,100.0,100.0,100.0
12.8,61.8,77.7,83.9,97.2
36.2,77.6,85.6,88.5,94.3
\end{filecontents*}
\pgfplotstabletypeset[%
				columns/n_keys_1/.style = {column name={1}, string type,},
				columns/n_keys_10/.style = {column name={10}, string type,},
				columns/n_keys_20/.style = {column name={20}, string type,},
				columns/n_keys_30/.style = {column name={30}, string type,},
				columns/n_keys_100/.style = {column name={100}, string type,},
                every head row/.style={after row=\hline},
                every last row/.style={before row=\hline \rowstyle{\bfseries}},
                color cells={min=-100,max=100,textcolor=black},
                /pgfplots/colormap={blackwhite}{rgb255=(165,225,165) color=(white) rgb255=(165,225,165)},
                /pgf/number format/fixed,
                /pgf/number format/precision=1,
                postproc cell content/.append style={/pgfplots/table/@cell content/.add={}{\%},},
                col sep=comma,]
                {accuracy_tables_data/single_prime/top_n_1.csv}}
            \\
            \hline
    \end{tabular}
    \end{adjustbox}
       \vspace{4mm}
\caption{Classification accuracy for single-prime features evaluated on TLS domain.}\label{table:singleprimetlsaccuracy}
\end{table}

The rest of this section is motivated by a setting when one wants to analyze a batch of correlated keys. Specifically, we assume a case of $k \geq 1$ keys $(p_1, q_1), \dots , (p_k, q_k)$ generated by the same source, where $p_1 = p_2 = \dots = p_k$. This scenario emerges in Section~\ref{sect:world_results}  and cannot be addressed by previously considered classifiers. If applied, the results would be drastically skewed since the classifier would consider each of $p_i$ separately, putting half of the weight on the shared prime. For that reason, we train a classifier that works on single primes rather than on complete private keys. Instead of feeding the classifier with a batch of $k$ private keys, we supply it with a  batch of $k+1$ unique primes from those keys. The selected features were modified accordingly: we extract the 5 most significant bits from the unique prime, its second least significant bit, and compute the ROCA and modular fingerprint for the single prime. We trained the classifier on the learning set limited to the TLS domain, as in Section~\ref{sect:tls_domain}.

On average, we achieve $28.8\%$ precision and $36.2\%$ recall when classifying a single prime. Table \ref{table:singleprimetlsaccuracy} shows the accuracy results in more detail. It should, however, be stressed that this classifier is meant to be used for batches of many keys at once. When considering a batch of $k \geq 10$ primes, the accuracy is more than $77$\%. The decrease in accuracy compared to Section \ref{sect:tls_domain} can be explained by the loss of information from the second prime. The features \textbf{mod} and \textbf{blum} are much less reliable when using only one prime. Since we can compute the most significant bits from a single prime at a time, we lost the information about the ordering of primes (since features \textbf{5p} and \textbf{5q} are correlated). These facts resulted in only nine separate groups of libraries being distinguishable. The following groups from the TLS domain are no longer mutually distinguishable: 5 and 13, 7 and 11, 8 and 9 and 10. 

\subsection{Methodology limitations}
The presented methodology has several limitations:

\textbf{Classification of an unseen source.}
Not all existing sources of RSA keys are present in our dataset for clustering analysis and classification. This means that attempting to classify a key from a source not considered in our study will bring unpredictable results. The new source may either populate some existing group or have a unique implementation, thus creating a new group. In both cases, the behaviour of the classifier is unpredictable. 

\textbf{Granularity of the classifier.}
There are multiple libraries in a single group. The user is therefore not shown the exact source of the key, but the whole group instead. This limitation has two main reasons: 1) Some sources share the same implementation and thus cannot be told apart. 2) The list of utilized features is narrow. There are infinitely many possible features in principle and some may hide valuable information that can further help the model performance. Nevertheless, the proposed methodology allows for an automatic evaluation of features using the naïve Bayes method which shall be considered in future work.

\textbf{Human factor.}
The clustering task in our study requires human knowledge. To be specific, the value of the threshold that splits the libraries into groups (for a particular feature) is established only semi-automatically. We manually confirmed the threshold -- when we could explain the difference between the libraries, or moved it otherwise. Summarized, this complicates the fully automatic evaluation on a large number of potential features. Once solved, the relative importance of the individual features could be measured.

\section{Real-world GCD-factorable keys origin investigation}
\label{sect:world_results}

Previous research \cite{lenstra2012,heninger2012mining,hastings2016weak,barbulescu2016rsa} demonstrated that a non-trivial fraction of RSA keys used on publicly reachable TLS servers is generated insecurely and is practically factorable. This is because the affected network devices were found to independently generate RSA keys that share a single prime or both primes. While an efficient factorization algorithm for RSA moduli is unknown, when two keys accidentally share one prime, the efficient factorization is possible using the Euclidean algorithm to find their GCD\footnote{Note that the keys sharing both primes are not susceptible to this attack but reveal their private keys to all other owners of the same RSA key pair.}. Still, the current number of public keys obtained from crawling TLS servers is too high to allow for the investigation of all possible pairs. However, the distributed GCD algorithm \cite{fastgcd} allows analyzing hundreds of millions of keys efficiently. Its performance was sufficient to analyze all keys collected from IPv4-wide TLS scans \cite{rapid7,censys_tls_ipv4} and resulted in almost 1\% of factorable keys in the scans collected at the beginning of the year 2016.     

After the detection of GCD-factorable keys, the question of their origin naturally followed. Previous research addressed it using two principal approaches: 1) an analysis of the information extractable from the certificates of GCD-factorable keys, and 2) matching specific properties of factored primes with primes generated by a suspected library -- OpenSSL. The first approach allowed to detect a range of network routers that seeded their PRNG shortly after boot without enough entropy, what caused them to occasionally generate a prime shared with another device. These routers contained a customized version of the OpenSSL library, what was confirmed with the second approach, since OpenSSL code intentionally avoids small factors of $p-1$ as shown by~\cite{Mironov}.

While this suite of routers was clearly the primary source of the GCD-factorable keys, are they the sole source of insecure keys? The paper \cite{hastings2016weak} identified 23 router/device vendors that used the code of OpenSSL (using specific OpenSSL fingerprint based on avoidance of small factors in $p-1$ and information extracted from the certificates). Eight other vendors (DrayRek, Fortinet, Huawei, Juniper, Kronos, Siemens, Xerox, and ZyXEL) produced keys without such OpenSSL fingerprint, and the underlying libraries remained unidentified. In the rest of this section, we build upon the prior work to identify probable sources of the GCD-factorable keys that \emph{do not} originate from the OpenSSL library.

Two assumptions must be met to employ the classifier studied in Section~\ref{sect:factorable_tls_domain}. First, we assume that \emph{when a batch of GCD-factored keys shares a prime, they were all generated by sources from a single classification group}. This conjecture is suggested in~\cite{hastings2016weak,heninger2012mining} and supported by the fact that when distinct libraries differ in their prime generation algorithm, they will produce different primes even when initialized from the same seed. On the other hand, when they share the same generation algorithm, they inevitably fall into the same classification group. Second, we assume that \emph{if the malformed keys share only single prime, the PRNG was reseeded with enough entropy before the second prime got generated}. This is suggested by the failure model studied for OpenSSL in~\cite{heninger2012mining} and implies that the second prime is generated as it normally would be. 

Leveraging these conjectures, the rest of this section tracks the libraries responsible for GCD-factorable keys while not relying on the information in the certificates. First, we describe the dataset gathering process, as well as the factorization of the RSA public keys. Later, successfully factored keys are analyzed, followed with a discussion of findings. 

\section{Datasets of GCD-factorable TLS keys}
\label{sect:batchgcd_dataset}

The input dataset with public RSA keys (both secure and vulnerable ones) was obtained from the Rapid7 archive. All scans between October 2013 and July 2019 (mostly in one or two weeks period) were downloaded and processed, resulting in slightly over 170 million certificates. Only public RSA keys were extracted, and duplicates removed, resulting in 112 million unique moduli. On this dataset, the \emph{fastgcd} \cite{fastgcd} tool based on \cite{bernstein2004find} was used to factorize the moduli into private keys. A detailed methodology of this procedure is discussed in Appendix~\ref{subsec:gcd_dataset}.

\subsection{Batching of GCD-factorable keys}
\label{sect:keys_batching}

Would the precision and recall of our classifier be $100$\%, one could process the factored keys one by one, establish their origin library and thus detect all sources of insecure keys. But since the classification accuracy of the single-prime TLS classifier\footnote{Note that without using single-prime model, the results are biased as the shared prime is considered multiple times in the classification process.} with a single key is only $36$\%, we apply three adjustments: 1) batch the GCD-factorable keys sharing the same prime (believed to be produced by the same library); 2) analyze only the batches with at least 10 keys (therefore with high expected accuracy); 3) limit the set of the libraries considered for classification only to the single-prime TLS domain. Since the keys from the OpenSSL library were already extensively analyzed by~\cite{hastings2016weak}, we use the \textbf{mod} feature to reliably mark and exclude them from further analysis. By doing so, we concentrate primarily on the non-OpenSSL keys that were not yet attributed.  The exact process for classification of  factored keys in batches is as follows:
\begin{enumerate}
    \item Factorize public keys from a target dataset (e.g., Rapid7) using \emph{fastgcd} tool.
    \item Form batches of factored keys that share a prime and assume that they originate from the same classification group. 
    \item Select only the batches with at least $k$ keys (e.g., 10).
    \item Separate batches of keys that all carry the OpenSSL fingerprint. As a control experiment, they should classify only to a group with the OpenSSL library.
    \item Separate batches without the OpenSSL fingerprint. This cluster contains yet unidentified libraries.    
    \item Classify the non-OpenSSL cluster using a single-prime TLS classifier. 
\end{enumerate}

\subsection{Source libraries detected in GCD-factorable TLS keys }

\begin{table}
    \centering
    \begin{tabular}{|p{6.3cm}|C{3cm}|}
        \hline
        Group(s) & \# batches  \\ [1mm]
        \hline
        1 (OpenSSL) & 2230 \\ 
        2 (8-bit OpenSSL) & 3 \\ 
        8 $|$ 9 $|$ 10 (various libraries, see Figure \ref{fig:tls_dendrogram}) & 278 \\ 
        3; 4; 6; 12; 5 $|$ 13; 7 $|$ 11 & 0 (\emph{improbable}) \\ 
        \hline
      \end{tabular}
      \vspace{4mm}
    \caption{Keys that share a prime factor belong to the same batch. Classification of most batches resulted in OpenSSL as the likely source. The rest of the batches were likely generated by libraries in the combined group 8 $|$ 9 $|$ 10.}
    \label{tab:batch_classif_tls}
\end{table}

In total, we analyzed more than 82 thousand primes divided into 2511 batches. While each batch has at least 10 keys in it, the median of the batch size is 15. Among the batches, $88.8$\% of them exhibit the OpenSSL fingerprint. This number well confirms the previous finding by~\cite{hastings2016weak} that also captured the OpenSSL-specific fingerprint in a similar fraction of keys.  We attribute three other batches as coming from the OpenSSL (8-bit fingerprint), an OpenSSL library compiled to test and avoid divisors of $p-1$ only up to 251. Importantly, slightly more than 11\% of batches were generated by some library from groups 8, 9, or 10, which are not mutually distinguishable when only a single prime is available. There are also negative results to report. With the accuracy over 80\% (for a batch size of 15) and no batches attributed to any of groups 3, 4, 6, 12, 5 $|$ 13, or 7 $|$ 11, it is very improbable that any GCD-factorable keys originate from the respective sources in these libraries. 

\section{Related work}
\label{sect:discusussion}
The fingerprinting of devices based on their physical characteristics, exposed interfaces, behaviour in non-standard or undefined situations, errors returned, and a wide range of various other side-channels is a well-researched area. The experience shows that finding a case of a non-standard behaviour is usually possible, while making a group of devices indistinguishable is very difficult due to an almost infinite number of observable characteristics, resulting in an arms race between the device manufacturers and fingerprinting observers.    

Having the device fingerprinted is helpful to better understand the complex ecosystem like quantifying the presence of interception middle-boxes on the internet \cite{durumeric2017security}, types of clients connected or version of the operating system. Differences may help point out subverted supply chains or counterfeit products. 

When applied to the study of cryptographic keys and cryptographic libraries, researchers devised a range of techniques to analyze the fraction of encrypted connections, the prevalence of particular cryptographic algorithms, the chosen key lengths or cipher suites \cite{durumeric2013analysis,sslObservatory10,barbulescu2016rsa,sshCiphers16,Gustafsson2017,cangialosi2016measurement,vandersloot2016towards}. 
Information about a particular key is frequently obtained from the metadata of its certificate. 

Periodical network scans allow to assess the impact of security flaws in practice. The population of OpenSSL servers with the Heartbleed vulnerability was measured and monitored by \cite{durumeric2014matter}, and real attempts to exploit the bug were surveyed. If the necessary information is coincidentally collected and archived, even a backward introspection of a vulnerability in time might be possible.

The simple test for the ROCA vulnerability in public RSA keys allowed to measure the fraction of citizens of Estonia who held an electronic ID supported by a vulnerable smartcard, by inspecting the public repository of eID certificates \cite{roca-2017-ccs-nemec}. 
The fingerprinting of keys from smartcards was used to detect that private keys were generated outside of the card and injected later into the eIDs, despite the requirement to have all keys generated on-card \cite{parsovs_usenix}.  

The attribution of the public RSA key to its origin library was analyzed by \cite{1mrsa}. Measurements on large datasets were presented in \cite{2017-acsac-nemec}, leading to accurate estimation of the fraction of cryptographic libraries used in large datasets like IPv4-wide TLS.  While both \cite{1mrsa} and \cite{2017-acsac-nemec} analyze the public keys, private keys can be also obtained under certain conditions of faulty random number generator \cite{lenstra2012,cryptosense,hastings2016weak,heninger2012mining,debianFlaw}. The origin of weak factorable keys needs to be identified in order to notify the maintainers of the code to fix underlying issues. A combination of key properties and values from certificates was used.

\section{Conclusions}
\label{sect:conclusions}

We provide what we believe is the first wide examination of properties of RSA keys with the goal of attribution of private key to its origin library. The attribution is applicable in multiple scenarios, e.g., to the analysis of GCD-factorable keys in the TLS domain. We investigated the properties of keys as generated by \numlibs cryptographic libraries, identified biased features in the primes produced, and compared three models based on Bayes classifiers for the private key attribution. 

The information available in private keys significantly increases the classification performance compared to the result achieved on public keys~\cite{1mrsa}. Our work enables to distinguish 26 groups of sources (compared to 13 on public keys) while increasing the accuracy more than twice w.r.t. random guessing. When 100 keys are available for the classification, the correct result is almost always provided ($>99\%$) for 19 out of 26 groups. 

Finally, we designed a method usable also for a dataset of keys where one prime is significantly correlated. Such primes are found in GCD-factorable TLS keys where one prime was generated with insufficient randomness and would introduce a high classification error in the unmodified method. As a result, we can identify libraries responsible for the production of these GCD-factorable keys, showing that only three groups are a relevant source of such keys. The accurate classification can be easily incorporated in forensic and audit tools. 

While the bias in the keys usually does not help with factorization, the cryptographic libraries should approach their key generation design with a great care, as strong bias can lead to weak keys~\cite{roca-2017-ccs-nemec}. We recommend to follow a key generation process with as little bias present as possible.

\textbf{Acknowledgements}
The authors would like to thank anonymous reviewers for their helpful comments. P. Svenda and V. Matyas were supported by Czech Science Foundation project GA20-03426S. Some of the tools used and other people involved were supported by the CyberSec4Europe Competence Network. Computational resources were supplied by the project e-INFRA LM2018140.

\bibliography{bibliography}

%

\appendix

\section{Detailed discussion of classifier results}~\label{subsec:discussion_results}

Some groups are accurately classified and rarely misclassified even with a single key available: namely group 1 (Infineon prior 2017, distinct because of the ROCA fingerprint), group 2 (Giesecke\&Devrient SmartCafe 4.x and 6.0), group 24 (standard OpenSSL \emph{without} the FIPS module enabled) and group 26 (Giesecke\&Devrient SmartCafe 7.0) are all classified with more than $96\%$ recall. Groups 1, 2, and 26 are rarely misclassified as origin library (false positive). The keys from group 25 (OpenSSL avoiding only 8-bit small factors in $p-1$) are misclassified as group 24 (standard OpenSSL) in $31.6\%$ cases, which still identifies the origin library correctly, only misidentifies the OpenSSL compile-time configuration. 

In contrast, keys from groups 7, 10, 11, 14, 15, and 17 are almost always misclassified (less than $8\%$ recall, some even less than 1\%). However, as discussed in the next section, if some additional information is available and can be considered, this misclassification can be largely remediated. 

Keys from group 7 (Libgcrypt) are mostly misclassified as group 6 (PGP SDK 4, $64.5\%$) or group 13 (Gemalto GXP E64, $20.2\%$). As libgcrypt is a commonly used library while groups 6 and 13 correspond to a very old library and card, this case demonstrates the possibility for further classifier improvement when some prior knowledge is available. E.g., for the TLS domain, groups corresponding to old smartcards or non-TLS libraries can be ruled out from the process. 

Group 10 (Bouncy Castle since 1.54, Mocana 7.x or HSM Thales nShieldF3) is misclassified as group 12 (smartcard Taisys SIMoME, $36.3\%$) or group 5 (Mocana 6.x $21.0\%$). Additional information can improve classification accuracy as the Taisys smartcard is unlikely source for the most usage domains. If Mocana library actually generated the key, only the identified version is incorrect. 

Group 11 (cryptlib, Safenet HSM Luna SA-1700, and Feitian and Oberthur cards) is misclassified as group 12 (smartcard Taisys, $50.2\%$) or group 20 (Oberthur Cosmo Dual, $20.4\%$). This is a very similar case as for group 10.

Group 14 (Microsoft and Crypto++, prevalent group) is misclassified as group 6 (PGP SDK 4, $23.9\%$), group 12 (card Taisys, $20.1\%$), group 13 (card Gemalto GXP E64, $13.5\%$) or group 5 (Mocana 6.x, $10.7\%$). Again, for the TLS domain, the only real misclassification problem is with the Mocana 6.x library.

Group 15 (large group with multiple frequently used libraries) is misclassified as group 12 (card Taisys, $27.2\%$), group 13 (card Gemalto GXP E64, $18.1\%$), group 20 (card Oberthur, $11.7\%$) or group 6 (PGP SDK 4, $32.3\%$). For the TLS domain, no group from the misclassified ones is likely. 

Group 17 (Nettle, Cryptix, FlexiProvider) is misclassified as multiple other groups where only groups 5 (Mocana 6.x) and 9 (Bouncy Castle prior 1.54 and SunRsaSign OpenJDK 1.8) cannot be ruled out as unlikely for the TLS domain.

\begin{table*}
    \centering
    \begin{adjustbox}{width=1\textwidth, center}
    \begin{tabular}{|@{}c@{}|@{}c@{}|@{}c@{}|@{}c@{}|}
		\hline
			& Top 1 match & Top 2 match & Top 3 match \\
			{\begin{filecontents*}{accuracy_tables_data/complex/groups.csv}
batch_keys
{Group 1}
{Group 2}
{Group 3}
{Group 4}
{Group 5}
{Group 6}
{Group 7}
{Group 8}
{Group 9}
{Group 10}
{Group 11}
{Group 12}
{Group 13}
{Group 14}
{Group 15}
{Group 16}
{Group 17}
{Group 18}
{Group 19}
{Group 20}
{Group 21}
{Group 22}
{Group 23}
{Group 24}
{Group 25}
{Group 26}
Average
\end{filecontents*}
\pgfplotstabletypeset[%
                every head row/.style={after row= \hline},
                /pgf/number format/fixed,
                /pgf/number format/precision=1,
                col sep=comma,
                every last row/.style={before row=\hline \rowstyle{\bfseries}},
                columns/batch_keys/.style={column name={\#keys in batch}, reset styles,string type,},]{accuracy_tables_data/complex/groups.csv}}
			& {\begin{filecontents*}{accuracy_tables_data/complex/top_n_1.csv}
n_keys_1,n_keys_2,n_keys_3,n_keys_5,n_keys_10
100.0,100.0,100.0,100.0,100.0
100.0,100.0,100.0,100.0,100.0
86.3,98.1,99.8,100.0,100.0
92.7,99.3,99.9,100.0,100.0
60.8,76.3,79.8,90.7,96.6
73.0,88.1,88.5,83.5,69.8
7.6,18.9,30.0,47.9,73.6
16.3,33.5,44.2,54.6,62.8
12.8,28.3,38.9,50.9,61.1
0.0,24.7,47.7,67.9,92.0
6.9,21.8,34.2,51.6,63.1
54.9,75.4,78.2,71.5,65.8
47.2,57.0,69.6,84.8,96.3
6.9,22.4,40.8,70.5,93.6
0.2,28.0,52.7,80.0,96.5
31.4,63.6,79.4,91.1,99.4
5.1,28.6,50.2,78.0,97.6
12.2,55.1,70.5,78.5,84.7
44.0,54.4,59.7,67.3,78.5
81.5,95.2,98.7,99.9,100.0
53.0,77.9,88.4,97.0,99.9
14.6,39.2,53.5,72.5,92.3
77.4,98.0,99.9,100.0,100.0
96.8,99.9,100.0,100.0,100.0
58.3,86.7,96.1,99.7,100.0
100.0,100.0,100.0,100.0,100.0
47.7,64.2,73.1,82.2,89.4
\end{filecontents*}
\pgfplotstabletypeset[%
				columns/n_keys_1/.style = {column name={1}, string type,},
				columns/n_keys_2/.style = {column name={2}, string type,},
				columns/n_keys_3/.style = {column name={3}, string type,},
				columns/n_keys_5/.style = {column name={5}, string type,},
				columns/n_keys_10/.style = {column name={10}, string type,},
                every head row/.style={after row=\hline},
                every last row/.style={before row=\hline \rowstyle{\bfseries}},
                color cells={min=-100,max=100,textcolor=black},
                /pgfplots/colormap={blackwhite}{rgb255=(165,225,165) color=(white) rgb255=(165,225,165)},
                /pgf/number format/fixed,
                /pgf/number format/precision=1,
                postproc cell content/.append style={/pgfplots/table/@cell content/.add={}{\%},},
                col sep=comma,]
                {accuracy_tables_data/complex/top_n_1.csv}}
			& {\begin{filecontents*}{accuracy_tables_data/complex/top_n_2.csv}
n_keys_1,n_keys_2,n_keys_3,n_keys_5,n_keys_10
100.0,100.0,100.0,100.0,100.0
100.0,100.0,100.0,100.0,100.0
98.2,100.0,100.0,100.0,100.0
94.8,99.7,100.0,100.0,100.0
71.5,90.1,93.6,98.7,99.9
92.8,92.8,97.7,98.2,99.9
77.3,95.5,98.8,99.9,100.0
27.5,56.2,73.5,91.3,99.2
37.7,65.7,79.1,90.4,99.0
18.4,44.1,60.8,79.8,96.1
56.7,87.2,95.9,99.4,100.0
72.2,85.0,95.4,98.1,100.0
52.9,68.6,80.9,93.8,99.5
7.7,41.0,69.7,90.8,99.3
2.5,43.4,65.4,90.2,99.4
40.9,70.6,85.4,96.5,100.0
18.3,51.2,71.9,92.0,99.7
45.2,91.0,98.2,100.0,100.0
54.5,88.3,97.3,99.9,100.0
97.2,100.0,100.0,100.0,100.0
95.2,99.7,100.0,100.0,100.0
78.0,98.2,99.8,100.0,100.0
96.8,99.9,100.0,100.0,100.0
100.0,100.0,100.0,100.0,100.0
87.6,97.9,99.6,100.0,100.0
100.0,100.0,100.0,100.0,100.0
66.3,83.3,90.9,96.9,99.7
\end{filecontents*}
\pgfplotstabletypeset[%
				columns/n_keys_1/.style = {column name={1}, string type,},
				columns/n_keys_2/.style = {column name={2}, string type,},
				columns/n_keys_3/.style = {column name={3}, string type,},
				columns/n_keys_5/.style = {column name={5}, string type,},
				columns/n_keys_10/.style = {column name={10}, string type,},
                every head row/.style={after row=\hline},
                every last row/.style={before row=\hline \rowstyle{\bfseries}},
                color cells={min=-100,max=100,textcolor=black},
                /pgfplots/colormap={blackwhite}{rgb255=(165,225,165) color=(white) rgb255=(165,225,165)},
                /pgf/number format/fixed,
                /pgf/number format/precision=1,
                postproc cell content/.append style={/pgfplots/table/@cell content/.add={}{\%},},
                col sep=comma,]
                {accuracy_tables_data/complex/top_n_2.csv}}
			& {\begin{filecontents*}{accuracy_tables_data/complex/top_n_3.csv}
n_keys_1,n_keys_2,n_keys_3,n_keys_5,n_keys_10
100.0,100.0,100.0,100.0,100.0
100.0,100.0,100.0,100.0,100.0
98.2,100.0,100.0,100.0,100.0
96.4,99.9,100.0,100.0,100.0
73.0,91.3,97.6,98.8,100.0
96.5,97.0,99.5,99.8,100.0
92.7,99.3,99.9,100.0,100.0
38.6,63.9,81.7,94.2,99.5
48.3,75.9,87.8,96.8,99.8
52.7,87.6,92.5,98.5,100.0
73.2,95.2,99.2,100.0,100.0
89.5,95.7,99.0,99.8,100.0
66.4,82.9,91.4,98.0,99.8
12.4,53.7,78.9,95.4,99.9
28.2,64.6,81.0,94.4,99.7
48.3,80.0,92.1,98.8,100.0
37.7,73.0,89.0,98.1,100.0
76.3,96.1,99.4,100.0,100.0
62.1,93.8,99.1,100.0,100.0
98.9,100.0,100.0,100.0,100.0
97.6,100.0,100.0,100.0,100.0
97.2,99.9,100.0,100.0,100.0
100.0,100.0,100.0,100.0,100.0
100.0,100.0,100.0,100.0,100.0
93.9,99.7,100.0,100.0,100.0
100.0,100.0,100.0,100.0,100.0
76.1,90.4,95.7,98.9,100.0
\end{filecontents*}
\pgfplotstabletypeset[%
				columns/n_keys_1/.style = {column name={1}, string type,},
				columns/n_keys_2/.style = {column name={2}, string type,},
				columns/n_keys_3/.style = {column name={3}, string type,},
				columns/n_keys_5/.style = {column name={5}, string type,},
				columns/n_keys_10/.style = {column name={10}, string type,},
                every head row/.style={after row=\hline},
                every last row/.style={before row=\hline \rowstyle{\bfseries}},
                color cells={min=-100,max=100,textcolor=black},
                /pgfplots/colormap={blackwhite}{rgb255=(165,225,165) color=(white) rgb255=(165,225,165)},
                /pgf/number format/fixed,
                /pgf/number format/precision=1,
                postproc cell content/.append style={/pgfplots/table/@cell content/.add={}{\%},},
                col sep=comma,]
                {accuracy_tables_data/complex/top_n_3.csv}}
            \\
            \hline
       \end{tabular}
       \end{adjustbox}
       \vspace{4mm}
\caption{The average classification accuracy of the best performing Bayes classifier. In the $i$-th column we consider a classifier successful if the true source of the key is among $i$ best guesses of our model. Similarly, for each of the 3 columns we evaluate the success rate when $1,2,3,5$ or $10$ keys from the same group are available.
}\label{table:bayes_classifier}
\end{table*}

\section{Obtaining dataset of GCD-factorable keys}~\label{subsec:gcd_dataset}

The \emph{fastgcd} \cite{fastgcd} tool based on \cite{bernstein2004find} was used to perform the search for the GCD-factorable keys. Only valid RSA keys were considered\footnote{The factorization occasionally finds small prime factors up to $2^{16}$, likely because the public key (certificate) was damaged, e.g., by a bit flip.}. Running the \emph{fastgcd} tool for a high number of keys (around 112 million for Rapid7 dataset) requires an extensive amount of RAM. Running the tool on a machine with 500 GB of RAM resulted in only a few factored keys, all sharing just tiny factors, while the tool did not produce any errors or warnings. The same computation on a subset of 10 million keys revealed a substantial number of large factors. 
Likely, the \emph{fastgcd} tool requires even more RAM for the correct functioning with such a large number of keys. 
To solve the problem, we partitioned the time-ordered dataset into two subsets of 50 and 62 million keys with an additional third subset with 50 million keys that partially overlapped both previous partitions. By doing so, we miss GCD-factorable keys that appeared in the dataset separated by a considerable time distance (2-3 years). We hypothesise that if a prevalent source starts producing GCD-factorable keys, we capture a sufficiently large batch of them within a single subset. In total, we have acquired 114 thousand unique factors from the whole dataset.

\end{document}